# Research on Pinches driven by SPEED 2 Generator: Hard X-ray and Neutron Emission in Plasma Focus Configuration.

Leopoldo Soto[1], José Moreno[1], Patricio Silva[1], Gustavo Sylvester[1], Marcelo Zambra[1], Cristian Pavez[1,2], Verónica Raspa[3], Fermín Castillo[4] and Walter Kies[5].

[1] *Comisión Chilena de Energía Nuclear, Casilla 188-D, Santiago, Chile*
[2] *Universidad de Concepción, Chile*
[3] *PLADEMA, CONICET and INFIP, Universidad de Buenos Aires, Argentina*
[4] *Insititnto de Ciencias Nucleares, UNAM, México*
[5] *Heinrich-Heine-Universität, Düsseldorf, Germany*

## Abstract

SPEED2 is a generator based on Marx technology and was designed in the University of Düsseldorf. SPEED2 consists on 40 +/- Marx modules connected in parallel (4.1 µF equivalent Marx generator capacity, 300 kV, 4 MA in short circuit, 187 kJ, 400 ns rise time, $dI/dt \sim 10^{13}$ A/s). Currently the SPEED2 is operating at the Comisión Chilena de Energía Nuclear, CCHEN, Chile, being the most powerful and energetic device for dense transient plasma in the Southern Hemisphere. Most of the previous works developed in SPEED2 at Düsseldorf were done in a plasma focus configuration for soft X-ray emission and the neutron emission from SPEED2 was not completely studied. The research program at CCHEN considers experiments in different pinch configurations (plasma focus, gas puffed plasma focus, gas embedded Z-pinch, wire arrays) at current of hundred of kiloamperes to mega-amperes, using the SPEED2 generator. The Chilean operation has begun implementing and developing diagnostics in a conventional plasma focus configuration operating in deuterium in order to characterize the neutron emission and the hard X-ray production. Silver activation counters, plastics CR39 and scintillator-photomultiplier detectors are used to characterize the neutron emission. Images of metallic plates with different thickness are obtained on commercial radiographic film, Agfa Curix ST-G2, in order to characterize an effective energy of the hard X-ray outside of the discharge.

## Introduction

SPEED2 is a generator based on Marx technology and was designed in the University of Düsseldorf. SPEED2 consists on 40 +/- Marx modules connected in parallel. Each module has 6 capacitors (50kV, 0.625 µF, 20nH) and 3 sparkgaps, so the pulse power generator SPEED2 is a medium energy and large current device (4.1 µF equivalent Marx generator capacity, 300 kV, 4 MA in short circuit, 187 kJ, 400 ns rise time, $dI/dt \sim 10^{13}$ A/s) [2]. The SPEED2 arrived at the Comisión Chilena de Energía Nuclear, CCHEN, in May 2001 from Düsseldorf University, Germany, and it is in operation since January 2002, being the most powerful and energetic device for dense transient plasma in the Southern Hemisphere. The research program at CCHEN considers experiments in different pinch configurations (plasma focus, gas puffed plasma focus, gas embedded Z-pinch, wire arrays) at current of hundred of kiloamperes to mega-amperes, using the SPEED2 generator [1].

In this paper a series of preliminary experimental results of the hard X-ray energy measurements in a plasma focus driven by SPEED2 are presented. Also some results related to neutron emission are shown. Operating in the plasma focus (PF) configuration, the pulse power generator SPEED2 have a maximum charging voltage of 300 kV [1], but for the actual measurements a 75 % of its 6 stage Marx generator was used to provide a charging voltage of 180 kV through the capacitor bank of 3.1 µF and to store an equivalent energy of 50.22 kJ. Using Deuterium like filling gas the working gas pressure was between 2.1 and 5 mbar.

Most of the previous works developed in SPEED2 at Düsseldorf were done in a plasma focus configuration for soft X-ray emission and the neutron emission and hard X-ray emission from SPEED2 were not completely studied.

The neutron emission studies in plasma focus configuration include a) the characterization of the total yield vs pressure and vs peak current, b) angular distribution and mechanisms of neutron emission (thermonuclear and beam target), and c) the possibility to enhance the thermonuclear component of the neutron yield against the beam target component. SPEED2 use a special insulator, quartz covered with alumina, and it requires several shots to perform an appropriate neutron emission whose neutron yield dispersion between shots be lower than 30 %. We have not enough shots with the same insulator in order to achieve the proper operating conditions. Preliminary results obtained at CCHEN show a neutron yield of the order of $10^{10}$ neutrons per shot, the maximum value obtained up to now at CCHEN is $2 \times 10^{10}$ neutrons per shot. In Düsseldorf, a neutron yield of the order of $10^{11}$-$10^{12}$ neutrons per shot was obtained [3]. Time resolution neutron detection and plastics CR39 nuclear track detectors in several angles are now being implemented. In the reference [5] it is reported time integrated measurements of the neutron emission from $-90^0$ to $90^0$ including 9 angles of view in two plasma focus devices. The neutron flux was measured with CR-39 nuclear track detectors covered with polyethylene. The results are consistent with an angular uniform plateau (isotropic emission) plus a shape peaked in the direction of the axis of the discharge (anisotropic emission). Using this analysis the authors obtain an interesting result, the 70% of the total neutron emission is isotropic and only the 30% is anisotropic, because to the axial emission is concentrated only in a small solid angle, thus its contribution to the total emission is lower in comparison to the isotropic emission. The value usually reported, as the ratio between only to point of measured, axial and radial, may be misleading.

In principle, X-rays are generated in the PF devices by Bremmstrahlung from the thermal electrons; by line emission from high Z ions (if present because they form the filling gas or either as impurities); and by high energy electron beams (hundreds of keV) colliding with the anode. The first two contributions are in the range of 1 to a few tens of keV (soft X-rays) while the last one lies in the range of hundred of keV (hard X-rays), which help to study them separately (fig. 1). In spite of several experimental and theoretical studies of the X-rays emitted by PF devices, their generation and temporal and spatial evolution need more clarification.

The image of metallic plates, of different thickness, in radiographic films was used like diagnostic tool. The statistical study of the digitised image intensities allow us to obtain a preliminary energy characterization of the emitted X-rays from the SPEED2. Considering the classical exponential radiation decay relation through the matter, $I(x)/I_0 = \exp(-k \cdot x)$, is possible to obtain the effective linear mass attenuation coefficient (k) when different gray shades of the digitised images are linked with the $I(x)/I_0$ rate. This method of image analysis, developed by Raspa [4], allows to obtain a correlation between k, through the mass absorption coefficient, and the X-ray energy [6].

**Experimental Setup and Results**

The scheme of the experimental setup is showed at figure 1. The plasma focus device has a cathode and anode length of ≤ 120 mm and 80 mm respectively, and the external and internal electrode radius of 110 mm y 54 mm respectively. The insulator length is 65 mm with an external radius of 60 mm. Commercial radiographic films, Agfa Curix ST-G2, in an AGFA CURIX Ortho Regular chassis with an intensified screen, were installed in the discharge axis (A and B in fig. 1), outside the discharge chamber. The A film was at 45.5 to 97.5 cm from the anode basis, meanwhile the B film was over the range of 100 and 156 cm from the anode top (fig. 1). A Lead target, 2.0 mm thickness, was installed in the anode basis in order to sensibility increase the efficiency of the colliding electrons the X-ray energy in the anode basis [2].

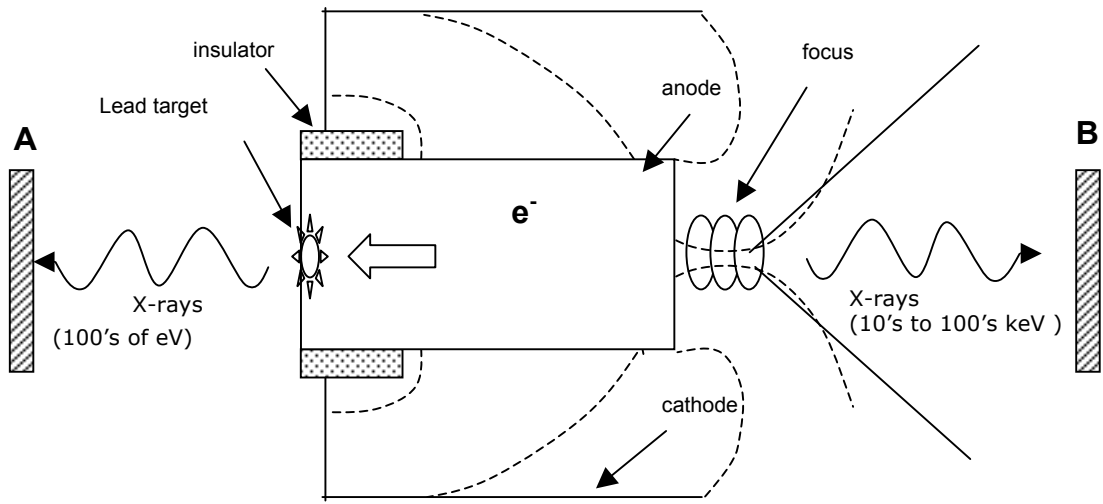

**Figure 1.** Scheme of the experimental device and X-ray axial emission.

Several metals were used like stepped filters and they impress its image on the films (A and B). The different elements used like filters are showed in the Table 1:

**Table 1.** Metallic elements used like stepped filters and its respective thickness

| Element | A Film Thickness (mm) | | | | | | | B Film Thickness (mm) | | | | | | |
|---|---|---|---|---|---|---|---|---|---|---|---|---|---|---|
| Pb (1) | | | | | | 6.0 | 8.0 | | | | | | 12.5 | 10.0 |
| Pb (2) | | | | | | 7.5 | 10.0 | | | | | 6.1 | 3.8 | 2.0 |
| Mo | | | 0.25 | 0.5 | 0.75 | 1.0 | 1.25 | | 0.25 | 0.5 | 0.75 | 1.0 | 1.25 | |
| Cu | 0.5 | 1.0 | 1.5 | 2.0 | 2.5 | 3.0 | 3.5 | 0.5 | 0.1 | 1.5 | 2.0 | 2.5 | 3.0 | |
| Ag | | 0.25 | 0.5 | 0.75 | 1.0 | 1.25 | 1.5 | | 0.25 | 0.05 | 0.75 | 1.0 | 1.25 | |
| Cd (1) | 0.5 | 1.0 | 1.5 | 2.0 | 2.5 | 3.0 | 3.5 | | 0.4 | 0.8 | 1.2 | 1.6 | 2.0 | |
| Cd (2) | | 0.43 | 0.88 | 1.3 | 1.74 | 2.18 | 2.61 | | 0.42 | 0.87 | 1.34 | 1.77 | 2.25 | |

(1) and (2) identify different arrangements of the same element

Figure 2 shows the typical digital computer image of a radiographic film obtained after an effective shot. An example of the intensity profile graphic of the different elements are showed at figure 3. The intensity profile graphic of the Lead (fig. 3a) allows to discriminate those images that are useful to determine the X-ray energy. In fact, the threshold of "non-irradiation" is verified by the Lead intensity level, that should be equal for the different existing arrangements (Pb(1) and Pb(2) in Table 1). However a good image contrast is also required for to determine the k values of the different involved metals. The adequate profile contrast is verified when a nice stepped curve is obtained for the intensity profile graphic of each element, an example is showed at figure 3b.

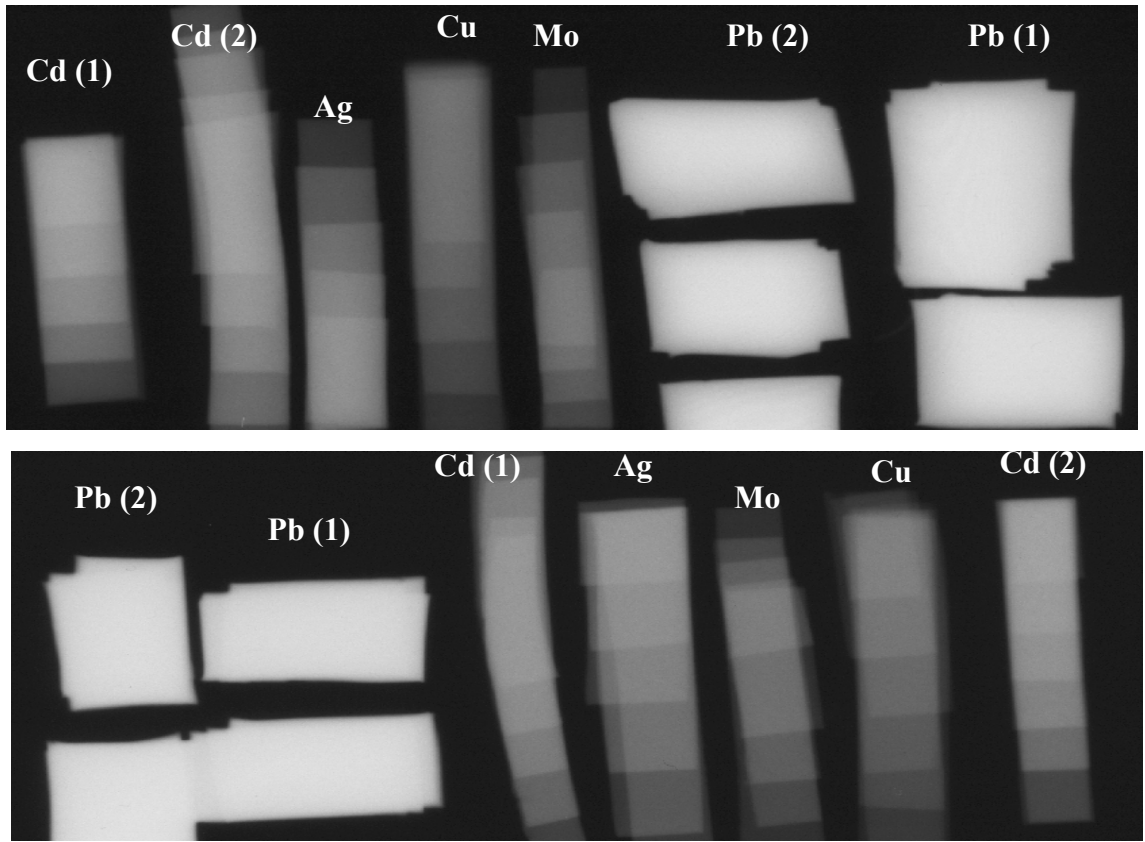

**Figura 2.** B and A radiographic images, top and bottom respectively, of the different metallic elements corresponding to axial x-ray emisión outside the discharge chamber.

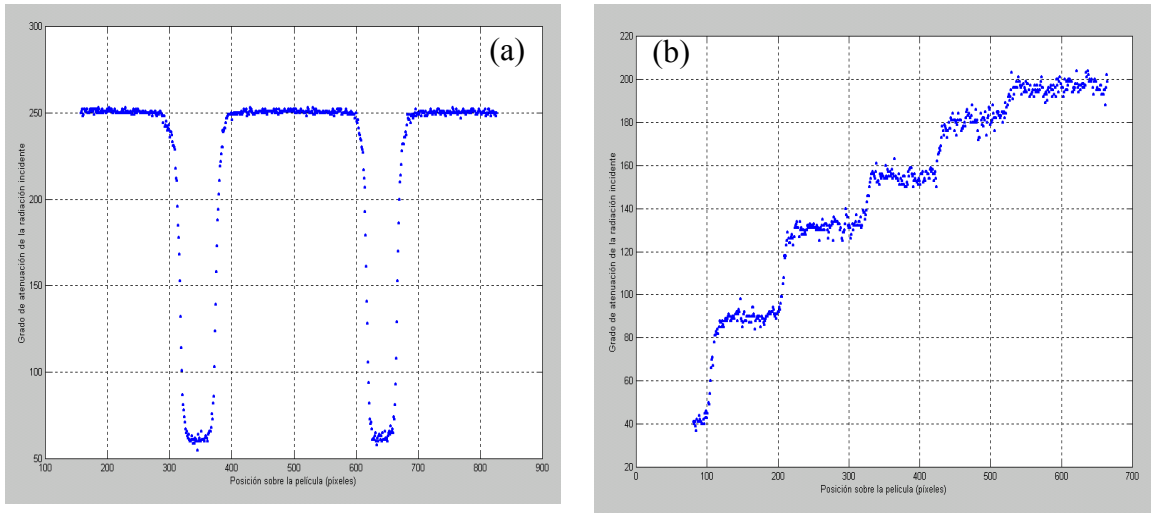

**Figure 3.** Intensity profile graphics: (a) Pb(2) intensity profile on the B film which verify the "non-irradiation level (reference level), and (b) example, in Ag, of the stepped profile which verify a good contrast.

The radiographic films were scanned (200 dpi resolution) and each time the Lead intensity profile graphic (fig. 3a) was obtained allowing to know its mean linear mass attenuation coefficient (k). The k value obtained is used like "non-irradiation" k reference for the corresponding film. The scan is over a region film that range between 240 and 550 pixels giving a percentage error ($\Delta k_i/k_i$) of 3.85 % in the best case and 7.15 % in the worst one.

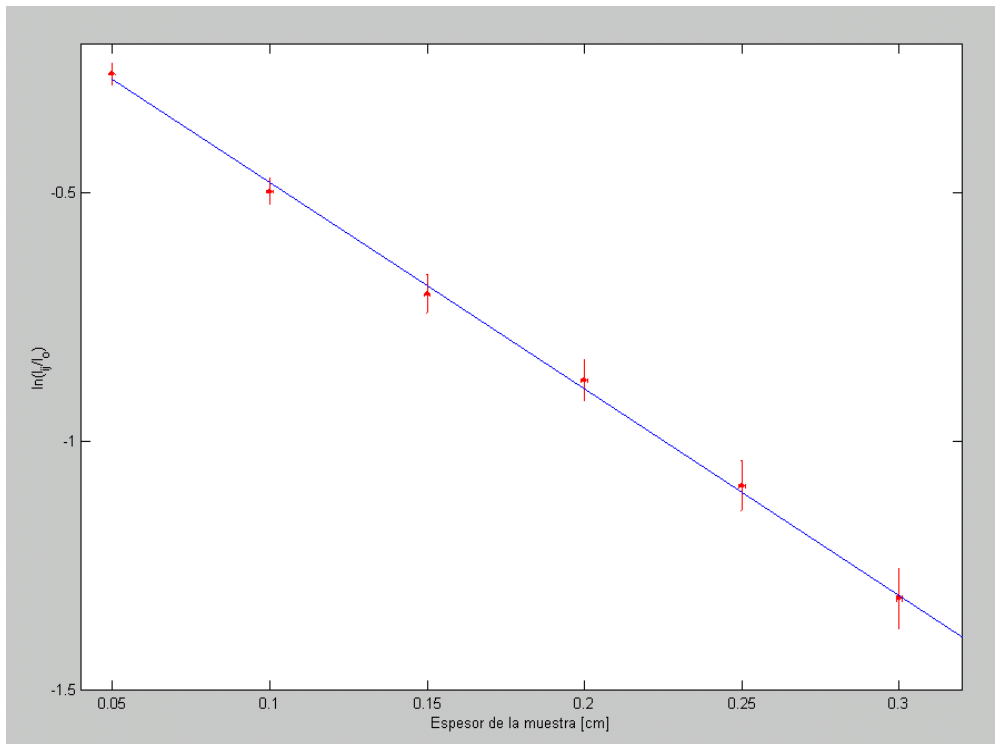

**Figure 4.** Intensity radiation rate against thickness of a specific element. Linear mass attenuation coefficient data points and the linked straight line whose slope give the mass attenuation coefficient.

The same procedure is used to obtain the linear mass attenuation coefficient against the thickness of the respective element. So, with this determined parameters is possible to obtain a graphic like is showed at figure 4. Each metallic element, images on the A and B films, have its proper $I(x)/I_0$ against its respective thickness. Also a linear fit through the data points allows to get, for each element, a k value but obtained this time from the slope with a percentage error ranged between 1.76 % and 15.39 %. So, at this stage we have a calculate mean value of k for each metallic element and a k value coming from the slope of the respective graphic like is showed at figure 4.

Relating the linear mass attenuation coefficient of each element with its respective mass attenuation coefficient [3], we obtain the value of the hard X-ray characteristic energy which impress the films after each shot. Table 2 briefly show the obtained results:

**Table 2.** Hard X-ray energy from the SPEED2

| Shot | Elem. | Z | Energy obtained from the slope (fig. 4) | | | | | | | Energy obtained from the data points (fig.4) | | | | | | |
|---|---|---|---|---|---|---|---|---|---|---|---|---|---|---|---|---|
| | | | k cm$^{-1}$ | Δk cm$^{-1}$ | Δk/k % | E keV | ΔE keV | ΔE/E % | <E> keV | k cm$^{-1}$ | Δk cm$^{-1}$ | Δk/k % | E keV | ΔE keV | ΔE/E % | <E> keV |
| 220704-01B | Cu | 28 | 2.84 | 0.19 | 6.69 | 123 | 3 | 2.44 | 124.2 ± 5.5 | 2.84 | 0.19 | 6.69 | 130 | 5 | 3.85 | 124.6 ± 7.6 |
| | Mo | 42 | 8.38 | 0.23 | 2.74 | 119 | 3 | 2.52 | | 8.38 | 0.23 | 2.74 | 120 | 2 | 1.67 | |
| | Ag | 47 | 10.72 | 0.36 | 3.36 | 124 | 2 | 1.61 | | 10.72 | 0.36 | 3.36 | 124 | 2 | 1.61 | |
| | Cd(2) | 48 | 9.12 | 0.52 | 5.70 | 127 | 2 | 1.57 | | 9.12 | 0.52 | 5.70 | 124 | 3 | 2.42 | |
| | Cd(1) | 48 | 9.06 | 0.6 | 6.62 | 128 | 2 | 1.56 | | 9.06 | 0.6 | 6.62 | 125 | 4 | 3.20 | |
| 220704-03B | Cu | 28 | 4.66 | 0.36 | 7.73 | 100 | 1 | 1.00 | 102.4 ± 2.2 | 4.66 | 0.36 | 7.73 | 96 | 3 | 3.13 | 98 ± 7.7 |
| | Mo | 42 | 13.4 | 1.14 | 8.51 | 99 | 1 | 1.01 | | 13.4 | 1.14 | 8.51 | 95 | 3 | 3.16 | |
| | Ag | 47 | 16.6 | 1.6 | 9.64 | 102 | 1 | 0.98 | | 16.6 | 1.6 | 9.64 | 98 | 3 | 3.06 | |
| | Cd(2) | 48 | 13.23 | 1.83 | 13.83 | 105 | 1 | 0.95 | | 13.23 | 1.83 | 13.83 | 100 | 4 | 4.00 | |
| | Cd(1) | 48 | 12.5 | 1.9 | 15.20 | 106 | 1 | 0.94 | | 12.5 | 1.9 | 15.20 | 101 | 4 | 3.96 | |
| 220704-05B | Cu | 28 | 6.1 | 0.06 | 0.98 | 85 | 2 | 2.35 | 86 ± 5.0 | 6.1 | 0.06 | 0.98 | 86 | 0.5 | 0.58 | 86.4 ± 2.1 |
| | Mo | 42 | 17.9 | 0.34 | 1.90 | 85 | 2 | 2.35 | | 17.9 | 0.34 | 1.90 | 85 | 1 | 1.18 | |
| | Ag | 47 | 24.2 | 0.75 | 3.10 | 85 | 2 | 2.35 | | 24.2 | 0.75 | 3.10 | 86 | 1 | 1.16 | |
| | Cd(2) | 48 | 19.9 | 0.35 | 1.76 | 87 | 2 | 2.30 | | 19.9 | 0.35 | 1.76 | 87 | 1 | 1.15 | |
| | Cd(1) | 48 | 19.7 | 0.52 | 2.64 | 88 | 3 | 3.41 | | 19.7 | 0.52 | 2.64 | 88 | 1 | 1.14 | |
| 230704-08B | Cu | 28 | - | - | - | - | - | - | 141 ± 4.0 | - | - | - | - | - | - | 141.3 ± 5.1 |
| | Mo | 42 | 5.42 | 0.41 | 7.56 | 138 | 2 | 1.45 | | 5.42 | 0.41 | 7.56 | 142 | 3 | 2.11 | |
| | Ag | 47 | 7.53 | 0.47 | 6.24 | 138 | 2 | 1.45 | | 7.53 | 0.47 | 6.24 | 141 | 2 | 1.42 | |
| | Cd(2) | 48 | 6.44 | 0.49 | 7.61 | 143 | 2 | 1.40 | | 6.44 | 0.49 | 7.61 | 140 | 3 | 2.14 | |
| | Cd(1) | 48 | 6.13 | 0.31 | 5.06 | 145 | 2 | 1.38 | | 6.13 | 0.31 | 5.06 | 142 | 2 | 1.41 | |
| 230704-08A | Cu | 28 | - | - | - | - | - | - | 143.3 ± 2.6 | - | - | - | - | - | - | 144 ± 4.8 |
| | Mo | 42 | 5.43 | 0.47 | 8.66 | 137 | 2 | 1.46 | | 5.43 | 0.47 | 8.66 | 142 | 3 | 2.11 | |
| | Ag | 47 | 7.25 | 0.46 | 6.34 | 141 | 1 | 0.71 | | 7.25 | 0.46 | 6.34 | 142 | 2 | 1.41 | |
| | Cd(2) | 48 | 5.5 | 0.25 | 4.55 | 146 | 1 | 0.68 | | 5.5 | 0.25 | 4.55 | 147 | 1 | 0.68 | |
| | Cd(1) | 48 | 5.66 | 0.53 | 9.36 | 149 | 1 | 0.67 | | 5.66 | 0.53 | 9.36 | 145 | 3 | 2.07 | |

The first column identify an individual shot, the film A or B can be identify because the last algorithm in the shot name is A or B. These selected shot are those that approved the intensity profile test early described (fig.3). The respective <E> columns give us the mean value of the X-ray energy for each shot. Only the shot 230704-08, the higher energy shot, was possible to analyse the film A and B; for the other ones it was impossible to analyse because a very bad contrast or a weak A (or a and B) film impression. It possible to observe, from the Table 2 in the case of the last shot, that it was impossible to determine the energy for the copper; surely this element was transparent for the ∪144 keV energy of X-ray which arrived on the films.

The energy of the hard X-rays, for each shot, was obtained independently analysing the impressed images. The energy dispersion has a reasonable value (ΔE/E ≤ 7.9 %) and a good agreement exists between the value obtained from the linear mass attenuation coefficient data points or from the slope in the $I(x)/I_0$ against thickness of the involved metallic element (Fig. 4). It can observe that the linear mass attenuation coefficient increase when the mass number increase [5], in our case an analyze of the Cadmium case is obliged.

## Acknowledgments

This work has been supported by the Fondecyt grants 1030062, 7040137, and the Chilean-Argentinean bilateral agrement (CCHEN-CNEA).

## References

1. G. Decker, W. Kies, M. Mälzig, C. van Calker and G. Ziethen "High Performance 300 kV Driver Speed 2 for MA Pinch Discharges": Nucl. Instr. And Meth. A249, 477-483 (1986)
2. "Research on Pinch Plasma Focus Devices of Hundred of Kilojoules to Tens of Joules", L. Soto, P. Silva, J. Moreno, G. Sylvester, M. Zambra, C. Pavez, L. Altamirano, H.Bruzzone, M. Barbaglia, Y. Sidelnikov, and Walter Kies, X Latin American Workshop on Plasma Physics, December 2003, to appears December 2004 in a Especial Issue of the Brazilian Journal in Physics
3. W. Kies, Private communication.
4. V. Raspa: "Estudio de un Equipo de Plasma Focus como Emisor de Rayos X de Alta Energía y su Aplicación a Radiografías no Convencionales"(in Spanish): Tésis de Licenciatura en Ciencias Físicas, Facultad de Ciencias Exactas y Naturales, Universidad de Buenos Aires, Argentina (2003)
5. F. Castillo, J. J. E. Herrera, J. Rangel, M. Milanese, R. Moroso, J. Pouzo, J. I. Golzarri, and G. Espinosa, Plasma Phys. and Control. Fusion **45**, 289 (2003).
6. J.H. Hubbell and S.M. Seltzer, "Tables of X-Ray Mass Attenuation Coefficients and Mass Energy-Absorption Coefficients (version 1.4)": [Online], available at http://physics.nist.gov/xaamdi [2004, September 7]. National Institute of Standards and Technology, Gaithersburg, MD.